\documentclass{appolb}
\usepackage{epsfig}

\usepackage{amsmath}
\usepackage{amsfonts}
\usepackage{amssymb}
\usepackage{psfrag}
\newcommand \mybf[1] {\mbox{\boldmath$ {\scriptstyle #1} $}}
\newcommand \Mybf[1] {\mbox{\boldmath$ {#1} $}}
\newcommand \vev [1] {\langle{#1}\rangle}
\newcommand\lr[1]{{\left({#1}\right)}}

\begin{document}

\title{Anomalous dimensions and reggeized gluon states%
\thanks{Presented at the 
Cracow School of Theoretical Physics, Zakopane, 2005}%
}
\author{Jan Kota{\'n}ski
\address{Jagiellonian University, ul. Reymonta 4, 30-059 Krak\'ow}
}
\maketitle
\begin{abstract}
Solving the BKP equation  
and comparing with the structure
function of hadron
for deep inelastic scattering processes
we are able to find a relation between
reggeized $N-$gluon states and anomalous dimensions
of QCD.
To this end we perform analytical continuation of the
Reggeon energy and compare exponents of 
two different twist-series expansions for the hadron structure function.
This makes possible to calculate the anomalous dimensions
and determine the twist related to them.
\end{abstract}
  
\section{Introduction}
During this talk  I would like to present the work
which was performed about one year ago
in collaboration with G. Korchemsky and  A. Manashov
\cite{Korchemsky:2003rc}.
This work describes the way one can calculate the anomalous dimensions of QCD 
making use of the reggeized gluon states, \ie Reggeons.
This approach was first used in 1980
by Jaroszewicz \cite{Jaroszewicz:1982gr}  
who calculated anomalous dimensions coming from 
$N=2$ Reggeon state corresponding to twist $n=2$.
Later, the cases for higher twists with $N=2$ were
computed  by Lipatov \cite{Lipatov:1985uk,deVega:2002im}.
In Ref. \cite{Korchemsky:2003rc} we performed calculation for
Reggeon states with $N>2$.

In order to present the calculation, firstly, I will explain
briefly what the reggeized gluons are. Next, I will describe
the way one can relate them to the anomalous dimensions in
QCD. Finally, I will show you our results.

\section{Reggeized gluons and Deep Inelastic Scattering}
Let us consider the deep inelastic scattering of 
hadron, $P(p)$, off virtual photon, $\gamma^{\ast}(q)$,
where the Bj\"orken $x=Q^2/2(p q)$ with
$Q^2=-q_{\mu}^2$ and $M^2=p_{\mu}^2$.
In the Regge limit of small $x$:
\begin{equation}
M^2 \ll Q^2 \ll s^2=(p+q)^2=Q^2(1-x)/x
\label{eq:Reglim}
\end{equation}
resuming appropriate Feynman diagrams
one can formulate an effective 
field theory in which compound states of gluons,
\ie reggeized gluons, play a role
of a new elementary field.

In the limit (\ref{eq:Reglim}) the leading contribution to the structure
function of hadron $F(x,Q^2)$ comes from the Reggeons with 
total angular momentum $j \to 1$. Thus, the structure 
function can be expanded in terms of the moments
\begin{equation}
\widetilde{F}(j,Q^2)
\equiv 
\int_{0}^{1} dx\, x^{j-2} F(x,Q^2)=
 \sum_{N=2}^{\infty} \bar\alpha_s^{N-2} \widetilde{F}_N(j,Q^2)
\label{eq:Fmom}
\end{equation}
with the strong coupling constant $\bar{\alpha}_s=\alpha_s N_c /\pi$, where
\begin{equation}
\widetilde{F}_N(j,Q^2)=\sum_{\mybf{q}} 
\frac1{j-1+\bar \alpha_s E_N(\Mybf{q})}
\beta_{\gamma^*}^{\,\mybf{q}}(Q)\,
 \beta_p^{\,\mybf{q}}(M)\,.
\label{eq:Fmom2}
\end{equation}
In the above formula the impact factors 
\begin{equation}
\beta^{\mybf{q}}_{\gamma^*}(Q^2)=\int d^2 z_0\, 
\vev{\Psi_{\gamma^*}|\Psi_{\mybf{q}}(\vec z_0)}, \qquad
\beta^{\mybf{q}}_P(M^2)=\int d^2 z_0\, \vev{\Psi_{\mybf{q}}(\vec z_0)|\Psi_P}
\label{eq:impf}
\end{equation}
are the overlaps between the Reggeon wave-function 
$\Psi_{\mybf{q}}(\vec z_1,\ldots,\vec z_N;\vec z_0)$ and 
the wave-functions of the scattering particles.
The parameters $\{\vec z_i\}_{i=1,\ldots,N}$ correspond to
transverse Reggeon coordinates and are integrated out in the scalar product 
(\ref{eq:impf})  defined in (\ref{eq:sp}).
In order to find the quantized values of 
$E_N(\Mybf{q})$ and $\Mybf{q}$
 one has to solve the Schr\"odinger-like equation 
\begin{equation}
{\cal H}_N 
\Psi_{\mybf{q}}\left( \left\{ \vec{z}_k \right\}\right)=
 E_N \Psi_{\mybf{q}}\left(\left\{ \vec{z}_k 
\right\}\right)
\label{eq:Schr}
\end{equation}
which was first formulated for $N=2$ Reggeons
by Balitsky, Fadin, Kuraev and Lipatov 
\cite{Fadin:1975cb,Kuraev:1977fs,Balitsky:1978ic}
and later generalized for $N\ge2$ Reggeon by Bartels,
Kwieci{\'n}ski, Prasza{\l}owicz and Jaroszewicz 
\cite{Bartels:1980pe,Kwiecinski:1980wb,Jaroszewicz:1980rw}.

It turns out that 
after performing the multi-colour limit \cite{'tHooft:1973jz}
Eq.~(\ref{eq:Schr}) corresponds
to the Schr\"odinger equation of the non-compact
Heisenberg $SL(2,\mathbb{C})$ spin chain magnet model
\cite{Korchemsky:2001nx,Derkachov:2001yn,Derkachov:2002wz} 
where the $E_N(\Mybf{q})$ plays a role
of the total energy. The system becomes
integrable with the complete set of
the integrals of motion $\Mybf{q}=(q_2,\bar q_2\ldots,q_N,\bar q_N)$
that are also
called conformal charges.
Introducing holomorphic and anti-holomorphic 
coordinates\footnote{variables from the anti-holomorphic sector are
denoted by barred characters} 
Eq.~(\ref{eq:Schr})
separates into two independent equations where
\begin{equation}
{\cal H}_N \sim 
\sum_{k=0}^{N-1} 
\left[
H(z_k, z_{k+1})+
H(\bar z_k, \bar z_{k+1}) 
\right]
\label{eq:HN}
\end{equation}
with $H(z_k, z_{k+1})$ defined in Ref.\ \cite{Derkachov:2001yn}.
The quantity $E_N(\Mybf{q})$ is also called the Reggeon energy.
The eigenvalue of the lowest conformal charge, $q_2$, may be parametrized by
\begin{equation}
q_2=-h(h-1)+N s(s-1)
\label{eq:q2}
\end{equation}
where in QCD $(s=0,\bar{s}=1)$ are the complex spins of Reggeons
and $(h,\bar{h})$ define a spin of $N$-Reggeon state
\begin{equation}
h=\frac{1+n_h}{2}+i \nu_h \qquad \bar h=\frac{1 - n_h}{2}+i \nu_h
\end{equation}
with $n_h \in {\mathbb Z} $ and  
$\nu_h \in {\mathbb R}$.
The Hamiltonian (\ref{eq:HN})
is invariant under $SL(2,\mathbb{C})$ group transformation
\begin{equation}
z_k  \rightarrow 
\frac{a z_k+b}{c z_k+d} 
\qquad
{\bar z}_k  \rightarrow 
\frac{\bar a \bar z_k+\bar b}{\bar c \bar z_k+\bar d} 
\qquad
a d - b c = 1
\quad 
\bar a \bar d-\bar b \bar c=1
\label{eq:zzbar}
\end{equation}
and its eigenstates transform as 
\begin{equation}
\Psi_q(\{\vec z\};\vec z_0) \to  (cz_0+d)^{2h}(\bar c \bar z_0+\bar
d)^{2\bar h} \prod_{k=1}^N (cz_k+d)^{2s}(\bar c \bar z_k+\bar d)^{2\bar
s}\Psi_{\mybf{q}}(\{\vec z\};\vec z_0)\,.
\label{eq:Psitr}
\end{equation}

In order to solve (\ref{eq:Schr})
we use an algorithm based on 
the $Q-$Baxter method \cite{Baxter}.
It is very interesting, however, 
complicated 
and the reader is referred to 
Refs.\ \cite{Korchemsky:2001nx,Derkachov:2001yn,
Derkachov:2002wz,Kotanski:2001iq} for details.

\section{Anomalous dimensions and twist series}
Due to the scaling symmetry of the Reggeon states (\ref{eq:Psitr}) 
one can calculate the dimensions of the impact factors as
\begin{equation}
\beta^{\mybf{q}}_{\gamma^*}(Q^2)= C_{\gamma^*}^{\mybf{q}} Q^{-1-2i\nu_h}
\,,\qquad
\beta^{\mybf{q}}_p(M^2)=C_{p}^{\mybf{q}}\, M^{-1+2i\nu_h}
\label{eq:dimif}
\end{equation}
where $C_{\gamma^*}^{\mybf{q}}$ and
$C_{p}^{\mybf{q}}$  are dimensionless constants.
Substituting (\ref{eq:dimif}) into (\ref{eq:Fmom2})
one obtains 
\begin{equation}
\widetilde F_N(j,Q^2)=\frac{1}{Q^2}\sum_{\mybf{\ell}}\sum_{n_h\ge 0 }
\int_{-\infty}^\infty d\nu_h
\frac{C_{\gamma^*}^{\mybf{q}}\,
C_{p^{\phantom{*}}}^{\mybf{q}}}{j-1+\bar \alpha_s
E_N(\Mybf{q})} \lr{\frac{M}{Q}}^{-1+2i\nu_h}
\label{eq:mom3}
\end{equation}
where integer $n_h$ and $\Mybf{\ell}=(\ell_1,\ldots,\ell_{2N-4})$
enumerate the quantized values of conformal charges 
$\Mybf{q}=\Mybf{q}(\nu_h;n_h,\Mybf{\ell})$ 
\cite{Derkachov:2002wz,Derkachov:2002pb}.
The integral over $\nu_h$ is calculated by performing analytical continuation
of  $E_N(\Mybf{q}(\nu_h))$ into the complex $\nu_h-$plane, closing the 
integral contour in infinity and summing the residua inside the contour 
at $\nu_h(j)$ defined by the condition  
\begin{equation}
j-1+\bar\alpha_s E_N(\Mybf{q}(\nu_h(j);n_h,\Mybf{\ell}))=0\,.
\label{eq:con1}
\end{equation}
Thus, the moment of the structure function for $j \to 1$ 
is given by
\begin{equation}
\widetilde F_N(j,Q^2)\sim 
\sum_{\rm{res}}
\frac{1}{Q^2} \left(\frac{M}{Q}\right)^{-1+2i\nu_h(j)}\,.
\label{eq:mom4}
\end{equation}
The above formula will help us to relate the reggeized gluon states to
the anomalous dimensions.

On the other hand the moments of $F(x,Q^2)$ can be expanded
in inverse powers of the hard scale $Q$, \ie in the twist series, as
\begin{equation}
{\widetilde F}(j,Q^2)=
\sum_{n=2,3,\ldots} \frac{1}{Q^n} 
\sum_{a} C_n^a(j,\alpha_s(Q^2))\, 
\vev{p\,|\mathcal{O}^a_{n,j}|p}
\label{eq:twser}
\end{equation}
which is also called operator product expansion (OPE).
The Wilson operators, $\mathcal{O}^a_{n,j}$,
satisfy
\begin{equation}
Q^2\frac{d}{d Q^2}
\vev{p\,|\mathcal{O}^a_{n,j}(0)|p}
=\gamma_n^a(j)\,\vev{p\,|\mathcal{O}^a_{n,j}(0)|p}
\label{eq:evol}
\end{equation}
where $a$ enumerates operators 
with the same twist and the
anomalous dimensions may be expanded as
\begin{equation}
\gamma_n^a(j)=\sum_{k=1}^\infty\gamma_{k,n}^a(j) (\alpha_s(Q^2)/\pi)^k\,.
\label{eq:gam}
\end{equation}
In the limit $j \to 1$
the moment $\widetilde F(j,Q^2)$ takes a form
\begin{equation}
\widetilde F(j,Q^2)=\frac{1}{Q^2} \sum_{n=2,3,\ldots}
\sum_{a} \tilde{C}_n^a(j,\alpha_s(Q^2))\ 
\left(\frac{M}{Q}\right)^{n-2-2\gamma_n^a(j)}\,.
\label{eq:mom5}
\end{equation}
Now we are ready to compare the exponents in (\ref{eq:twser}) and
(\ref{eq:mom5}) that results in
\begin{equation}
\gamma_n(j)=(n-1)/2-i\nu_h(j)=[n-(h(j)+\bar h(j))]/2\,.
\label{eq:con2}
\end{equation}
Combining (\ref{eq:con1}) with (\ref{eq:con2}) using
the property that $\gamma_n(j)\to 0$ for $\bar \alpha_s \to 0$
we are able to compute the coefficients
in Eq.~(\ref{eq:gam}) and to determine the 
corresponding twist.

Thus making use of the above equations we can
extract $\gamma_n(j)$ from 
the expansion of $E_N(\Mybf{q}(\nu_h(j)))$ in the vicinity
of its poles:
\begin{equation}
E_N(\Mybf{q})=
-\left[{\frac{c_{-1}}{\epsilon}+c_0+c_1\,\epsilon+\ldots}\right]
\label{eq:Eser}
\end{equation}
at $i \nu_h=  i \nu_h^{\rm pole} + \epsilon$. Inverting Eq.~(\ref{eq:Eser})
and using Eq.~(\ref{eq:con2})
one obtains
\begin{equation}
\gamma_n(j)=-c_{{-1}}\left[\frac{\bar\alpha_s}{j-1}+c_{{0}}\,
\lr{\frac{\bar\alpha_s}{j-1}}^2+\left( c_{{1}}c_{{-1}}+c_{0}^{\,2}
\right)\lr{\frac{\bar\alpha_s}{j-1}}^3+\ldots\right]
\label{eq:gam2}
\end{equation}
where the coefficient $c_k=c_k(n,n_h,\Mybf{\ell})$ are defined by
(\ref{eq:Eser}).
Moreover, it turns out that the position of the energy poles:
\begin{equation}
E_N(\Mybf{q})\sim 
\frac{\gamma_{n}^{(0)}}{i\nu_h-(n-1)/2}
\label{eq:}
\end{equation}
determines the twist $n$:
\begin{equation} 
i\nu_h=(n-1)/2 
\quad
\mbox{with}
\quad 
n\ge N+n_h.
\label{eq:tw}
\end{equation}

\section{Results}

\subsection{Analytical Continuation}
After performing the analytical continuation of $E_N(\Mybf{q}(\nu_h))$
in complex $\nu_h-$space  
the Reggeon wave functions $\Psi_{\Mybf{q}}(\{\vec z_i\};\vec z_0)$
is no more normalizable with respect to the scalar product
\begin{equation}
\vev{\Psi_{\mybf{q}}(\vec z_0)|\Psi_{\mybf{q}'}(\vec z_0')}\equiv\int
\prod_{k=1}^N d^2 z_k 
\Psi_{\mybf{q}}(\{\vec z\};\vec z_0) \lr{\Psi_{\mybf{q}'}(\{\vec z\};\vec
z_0')}^*= \delta^{(2)}(z_0-z_0')\,\delta_{\mybf{q}\mybf{q}'}\,.
 \label{eq:sp}
\end{equation}
Moreover, the quantization conditions for $\Mybf{q}$ become relaxed, so that
\begin{equation}
\bar q_k \ne q^*_k  
\quad
\mbox{and}
\quad
\bar h \ne 1-h^*\,.
\label{eq:qqbar}
\end{equation}

\subsection{Two-Reggeon states}
For $N=2$ Reggeon states the energy 
\cite{Fadin:1975cb,Kuraev:1977fs,Balitsky:1978ic}
in known analytically
\begin{equation}
E_2(\nu_h,n_h)=\psi\lr{\frac{1+|n_h|}2+i\nu_h}
+\psi\lr{\frac{1+|n_h|}2-i\nu_h}-2\psi(1)\,,
\label{eq:E2}
\end{equation}
with  $\Psi(x)=\frac{d}{dx}\Gamma(x)$. It is 
an analytical function on the complex $\nu_h-$plane 
without the branching points and
with poles at $i\nu_h=\pm(n-1)/2$.
The leading twist $n=2$, which agrees with (\ref{eq:tw}),
corresponds to the pole at $i\nu_h=1/2$, while 
the anomalous dimension \cite{Jaroszewicz:1982gr}
\begin{equation}
\gamma_2(j)=\frac{\bar\alpha_s}{j-1}
  +2\zeta(3)\lr{\frac{\bar\alpha_s}{j-1}}^4
+2\zeta(5)\lr{\frac{\bar\alpha_s}{j-1}}^6 +\mathcal{O}(\bar\alpha_s^8)\,.
\label{eq:g2}
\end{equation}

\subsection{N-Reggeon states}
For more than $N=2$ Reggeons the energy $E_N$
is a multi-valued function with the branching point
in the complex $\nu_h-$plane where the cuts take a form 
\begin{equation}
E_N^{\pm} \sim a_k \pm b_k \sqrt{\nu_{{\rm br},k}-\nu_h}\,.
\label{eq:Ecuts}
\end{equation}
Poles at $i\nu_h=(n-1)/2$ with twist $n\ge N+n_h$.
We do not have a unique analytical formula so we 
evaluate $E_N(\Mybf{q})$ numerically \cite{Derkachov:2002wz}.

\subsection{$N=3$ Reggeon states with $n_h=0$}
\begin{figure}[th]
\psfrag{E_N/4}[cc][cc]{$E_3$} \psfrag{Im_nu_h}[cc][bc]{$\mbox{\large$i\nu_h$}$}
\psfrag{Re_nu_h}[cc][bc]{$\mbox{\large$\nu_h$}$}
\vspace*{3mm} 
\centerline{{\epsfysize4.5cm \epsfbox{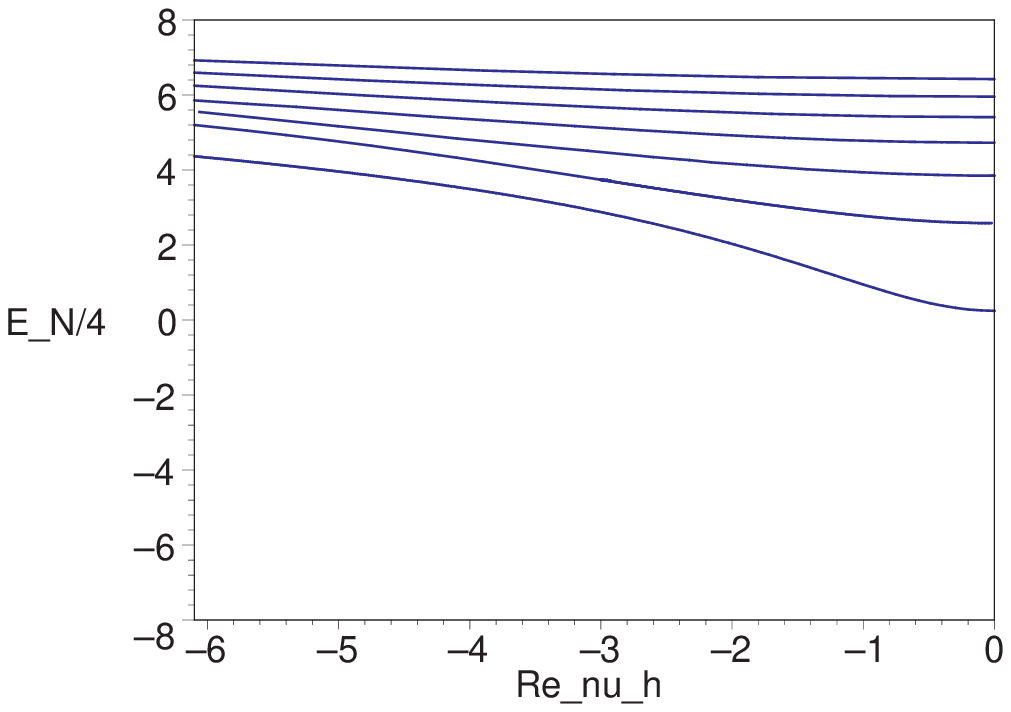}}{\epsfysize4.5cm
\epsfbox{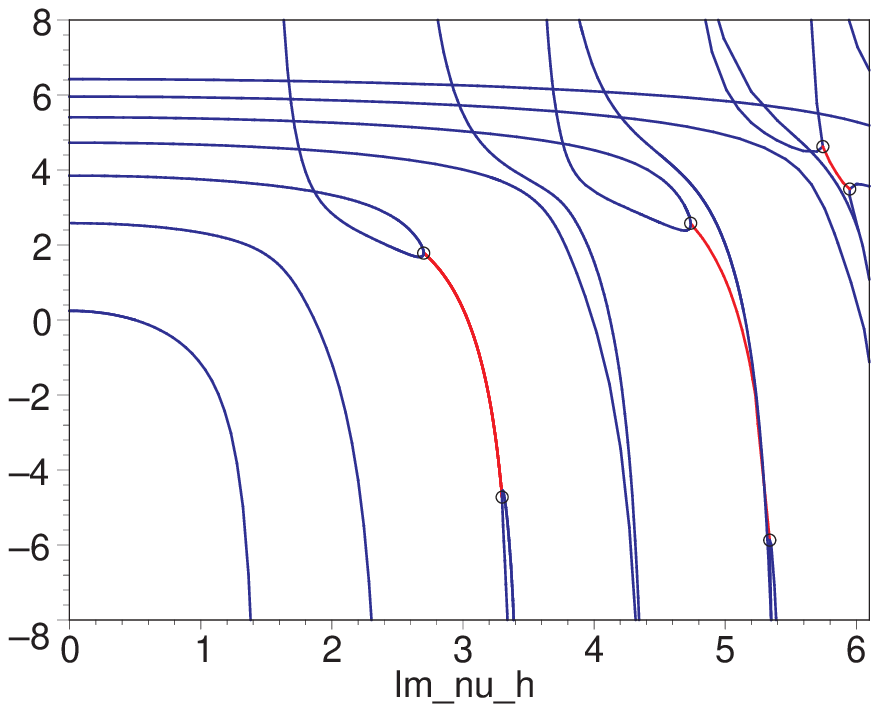}}}
%
%
%
\caption{The energy spectrum  of the $N=3$ Reggeon states $E_3(\nu_h;n_h,\Mybf{\ell})$
for $n_h=0$ and $\Mybf{\ell}=(0,\ell_2)$, with $\ell_2=2,\,4,\,\ldots,14$ from
the bottom to the top (on the left). Analytical continuation of the energy along
the imaginary $\nu_h-$axis  (on the right). The branching points are indicated by
open circles. The lines connecting the branching points represent 
$\mbox{Re} E_3(i \nu_h)$ \cite{Korchemsky:2003rc}.}
\label{fig:spec}
\end{figure}

Let us consider the case for $N=3$ and
$n_h=0$ where $\Mybf{\ell}=(0,\ell_2)$
with $\ell_2=2,\,4,\,\ldots,14$
what gives a condition $\bar q_3 + q_3 = 0$.
The spectral surfaces $E_3(\Mybf{q}(\nu_h))$
in this case along real and imaginary axes are shown in 
{Fig.~\ref{fig:spec}}.
On the left panel the energy
$E_3(\nu_h)$ is a monotonic function of real $\nu_h$.
However, on the right panel, \ie for imaginary values of $\nu_h$, 
branching points, denoted by open circles, appear.  
They glue together surfaces with the same quantum numbers.
They appear not only for purely imaginary $\nu_h$
but also for complex $\nu_h$.
However, one 
can notice that contribution to the structure function from the cuts
cancel each other so the (OPE) expansion in not broken.
The poles of $E_N(\nu_h)$ are localized at $i \nu^{\rm pole}=(n-1)/2$
which gives possible values of the twists
$n=4,6,8,\ldots$.

The leading twist $n=4$ in this sector comes from the poles
at $i \nu_h=3/2$ where 
\begin{equation}
E_3(3/2+\epsilon) =  
{\epsilon}^{-1} + \frac12 -\frac12\ \epsilon + 1.7021\,
\epsilon^2+\ldots
\label{eq:E34}
\end{equation}
which gives the anomalous dimension 
\begin{equation*}
\gamma_4^{(N=3)}(j) =
\frac{\bar\alpha_s}{j-1}-\frac12\lr{\frac{\bar\alpha_s}{j-1}}^2
-\frac14\lr{\frac{\bar\alpha_s}{j-1}}^3-1.0771
\lr{\frac{\bar\alpha_s}{j-1}}^4+\ldots\,.
\label{eq:gam34}
\end{equation*}

The energy around the other poles reads as follows 
\begin{eqnarray}
E_3(3/2+\epsilon) = \frac{1}{\epsilon} + \frac12 -\frac12\ \epsilon + 1.7021\,
\epsilon^2+\ldots \,, \nonumber \\
E_3(5/2+\epsilon) =  \frac{2}{\epsilon} + {\frac {15}{8}} 
- 1.6172\ \epsilon + 0.719\ {\epsilon}^{2}+\ldots \,, \nonumber  \\
E_3^{\rm(a)}(7/2+\epsilon)= \frac{1}{\epsilon} + \frac{11}{12} - 0.6806\
\epsilon - 1.966\ \epsilon^2+\ldots \,, \nonumber  \\
E_3^{\rm(b)}(7/2+\epsilon)= \frac{2}{\epsilon}
+\frac{15}4-3.2187\ \epsilon + 3.430\ \epsilon^2+\ldots \,, \nonumber  \\
E_3^{\rm(a)}(9/2+\epsilon)= \frac{2}{\epsilon} + \frac{125}{48} - 2.0687\
\epsilon + 1.047\ \epsilon^2+\ldots \,, \nonumber  \\
E_3^{\rm(b)}(9/2+\epsilon)= \frac{2}{\epsilon} + \frac{53}{12} - 2.4225\
\epsilon + 0.247\ {\epsilon}^{2}+\ldots
\label{eq:Elast}
\end{eqnarray}
and can be generally cast in the form:
\begin{equation}
E_3(i\nu_h^{\rm pole}+\epsilon) = 
\frac{R}{\epsilon} +2 {\cal E}(i\nu_h)+\mathcal{O}(\epsilon)
\label{eq:E3pat}
\end{equation}
where $R=2$ (or $R=1$ for  {\it even} $h=\frac{1}{2}+i \nu_h$)
and ${\cal E}(h)$ 
is energy of the Heisenberg model with $SL(2,{\mathbb R})$  
spin.
Making use of (\ref{eq:Eser}) and (\ref{eq:gam2})
one can calculate expansion coefficients of anomalous dimensions
corresponding to (\ref{eq:Elast}).

\subsection{$N=3$ descendent states with $n_h=1$}

Another interesting case for $N=3$ Reggeon states 
is when  $q_3=\bar q_3=0$ and $n_h=1$.
Such states  are called descendent 
of $N=2$ states because they 
possess the same quantum numbers as $N=2$ Reggeon states
and the energy
\begin{equation}
E_3^{\rm desc}(q_3=0,q_2;n_h=1)=E_2(q_2;n_h=1)
\label{eq:E3des}
\end{equation}
as a function of $\nu_h$
does not have branching points and its poles are
situated at $i \nu_h=1,2,3,\ldots$ with twists $n=3,5,7,\ldots$ .
According to (\ref{eq:tw})
this case is related to $N=2$ case which gives
one more argument
to call them descendent.

The leading twist, \ie $n=3$, corresponds to the energy pole where 
\begin{equation}
E_{3,\rm d}(1+\epsilon)=
{\epsilon}^{-1}+1-\epsilon -\left( 2\zeta(3) -1 \right)
 {\epsilon}^{2}+\ldots\,,
\label{eq:E3d}
\end{equation}
so that in this case the anomalous dimension
\begin{equation}
\gamma_3^{(N=3)}(j) = \frac{\bar\alpha_s}{j-1}-\lr{\frac{\bar\alpha_s}{j-1}}^2
+(2\zeta(3)+1)\lr{\frac{\bar\alpha_s}{j-1}}^4+\ldots\,.
\label{eq:gam3d}
\end{equation}

\subsection{Leading twist for higher $N$}

For higher $N$ the leading twist $n=N$.
However, considering {\it even} and
{\it odd} $N$'s separately one can notice that the leading twist 
comes from completely
different sectors.
For {\it even} $N$  it corresponds to 
the pole at $i \nu_h=(N-1)/2$
localized in the
sector where $n_h=0$. The energy pole residuum equals
to $(N-2)$ that gives anomalous dimensions
\begin{equation}
\gamma_N^{(N)}(j)=(N-2)\frac{\bar\alpha_s}{j-1}+ 
\mathcal{O}(\bar\alpha_s^2)
\label{eq:gamN}
\end{equation}
with the twist $n=N$.

For {\it odd} $N$'s in the sector with $n_h=0$
the minimal twist $n_{\rm min}=(N+1)$ corresponds
to the energy pole at $i \nu_h=N/2$.
For example, for $N=5$ we have
\begin{equation}
E_5(5/2+\epsilon)= \frac3{\epsilon} +  \frac{7}{6}
 +\ldots\,.
\label{eq:E5}
\end{equation}
However,
the real leading twist comes from the sector of descendent states
with $n_h=1$ and it corresponds to the pole at $i \nu_h=(N-1)/2$, \eg
\begin{equation}
E_{5,\rm d}(2+\epsilon)=
\frac{3+\sqrt{5}}{2\epsilon} + 1.36180 +\ldots\,.
\label{eq:E5d}
\end{equation}
As one can see, the pole residuum in this case has a more complicated form.

In each case the energy pole 
which gives the leading twist is situated on the same surface
as the state with the minimal energy $E_N(\nu_h)$
where $\nu_h\in \mathbb{R}$.

\section{Summary}
In this work following Ref.\ \cite{Korchemsky:2003rc} 
we have considered the deep inelastic scattering processes
of hadrons described by means of the reggeized gluon states.
Expanding the structure function (\ref{eq:Fmom}) we 
have used two approaches.
Firstly, we have expanded 
the structure function (\ref{eq:Fmom}) making use of the reggeized
gluon states which was possible due to  
the analytical continuation of the Reggeon
energy $E_N(\nu_h)$ into complex $\nu_h-$plane. 
Secondly, we have performed (OPE) expansion 
(\ref{eq:twser})
with evolution equation (\ref{eq:evol}), 
which defines the anomalous dimensions
of QCD.
Comparing exponents of these two series we have found
relation between $\gamma_n^a(j)$ and 
$N-$Reggeon states.
Since we are able to solve numerically
the BKP equation \cite{Derkachov:2001yn,Derkachov:2002wz} for
$N-$Reggeon states with $N\ge 2$,
we have calculated anomalous dimensions
and the twist coming from these states.

Contrary to $N=2$ Reggeon case
for $N\ge 3$ the energy $E_N(\nu_h)$
is a multi-valued function of complex $\nu_h$
parameter defined on complex Riemann
surface with the infinite number of branching points
that glue the surfaces with the same quantum numbers.
However, similarly to $N=2$ case,
the energy has poles only for purely imaginary $\nu_h$
and their position defines the
possible values of the twist (\ref{eq:tw}).
Fitting expansion coefficients of the energy
in the vicinity of these poles we have
calculated the anomalous dimensions of QCD.

It turns out that the leading twist $n$
coming from the $N-$Reggeon states  is equal to  
a number of reggeized gluons $N$.
However, we have to consider 
the states with {\it even} $N$
and the states
with {\it odd} $N$ separately.
For {\it even} $N$ the leading twist comes from the
sector where $n_h=0$, whereas for {\it odd} $N$
the minimal twist comes from the sector of descendent states
with $n_h=1$.

To sum up I would like to notice that
contrary to common approaches \cite{Levin:1992mu,Bartels:1993ih,Bartels:1993ke}
the present work goes beyond the leading
order of the twist expansion of the structure function
and describes contributions with non-leading
asymptotics that correspond to the higher
twists.

I would like to warmly thank to
 G.P. Korchemsky,
A.N.Manashov and S.\'E. Derkachov
with whom I have done this work. 
I am also grateful to 
M. Prasza{\l}owicz
and J. Wosiek 
for fruitful discussions.
This work was supported by
grants 
KBN-PB-2-P03B-43-24 and
KBN-PB-0349-P03-2004-27.


\end{document}